\documentstyle[epsfig]{mn}
\begin{document}
\LARGE
\normalsize

\title[Cygnus X-3]
{Infrared spectroscopic variability of Cygnus X-3 in outburst and quiescence}
\author[R.~P.~Fender, M.~M.~Hanson and G.~G.~Pooley]
{R. P. Fender,$^1$
\thanks{EC Marie Curie Fellow, email : rpf@astro.uva.nl}
M. M. Hanson,$^{2,3}$\thanks{Hubble Fellow, email : hanson@physics.uc.edu}
G. G. Pooley$^4$\thanks{email : ggp1@cam.ac.uk}\\
$^1$ Astronomical Institute `Anton Pannekoek', University of Amsterdam,
and Center for High Energy Astrophysics, Kruislaan 403, \\
1098 SJ, Amsterdam, The Netherlands\\
$^2$ Steward Observatory, University of Arizona, Tucson, AZ 85721, USA \\
$^3$ Department of Physics, University of Cincinnati, Cincinnati, OH 45221-0011, USA \\
$^4$ Mullard Radio Astronomy Observatory, Cavendish Laboratory, 
Madingley Road, Cambridge CB3 OHE\\
}

\maketitle

\begin{abstract}

We present four epochs of high-resolution infrared spectroscopy of the
peculiar X-ray binary Cygnus X-3. The observations cover quiescent,
small flaring and outburst states of the system as defined by radio
and X-ray monitoring. The underlying infrared spectrum of the source,
as observed during radio and X-ray quiescence and small flaring
states, is one of broad, weak He~{\sc ii} and N~{\sc v}
emission. Spectral variability in this state is dominated by
modulation at the 4.8 hr orbital period of the system. H-band spectra
confirm the significant hydrogen depletion of the mass donor. The
closest spectral match to the quiescent infrared spectrum of Cyg X-3
is an early-type WN Wolf-Rayet star.

In outburst, the infrared spectrum is dramatically different, with the
appearance of very strong twin-peaked He~{\sc i} emission displaying
both day-to-day variability and V(iolet)/R(ed) variations with orbital
phase.  We argue that the twin-peaked emission cannot arise in
relativistic jets or, unless the distance to Cyg X-3 is severely
overestimated, an accretion disc. The most likely explanation appears
to be an enhanced stellar wind from the companion.  Thus X-ray and
radio outbursts in this system are likely to originate in
mass-transfer, and not disc, instabilities, and the lengthening of the
orbital period will not be smooth but will be accelerated during these
outbursts.  Furthermore, the appearance of these lines is suggestive
of an asymmetric emitting region. We propose that the wind in Cyg X-3
is significantly flattened in the plane of the binary orbit. This may
explain the observed twin-peaked He~{\sc i} features as well as
reconciling a massive Wolf-Rayet secondary with the relatively small
optical depth to X-rays, if the disc wind is inclined at some angle to
the line of sight.  A small set of observations following outburst,
when the system was returning to a more quiescent X-ray and radio state,
reveal strong He~{\sc i} 2.058 $\mu$m absorption with a clear P-Cygni
profile, at the same time as the more common weak He~{\sc ii} and
N~{\sc v} features. In a disc-wind geometry this can be interpreted as
absorption in the densest, accelerating regions of the wind which can
be viewed directly if the disc is inclined at some angle to the line
of sight.

\end{abstract}


\section{Introduction}

Cygnus X-3 is a heavily obscured luminous X-ray binary in the Galactic
plane which displays a unique and poorly-understood combination of
observational properties. These include strong radio emission, with a
flat spectrum extending to (at least) mm wavelengths in quiescence
(e.g. Waltman et al. 1994; Fender et al. 1995) and giant flares which
are associated with a relativistic jet (e.g. Geldzahler et al. 1983;
Fender et al. 1997; Mioduszewski et al. 1998).  In the infrared the
system is bright with occasional rapid flare events and thermal
continuum consistent with a strong stellar wind (e.g. van Kerkwijk et
al. 1996; Fender et al. 1996). There is no optical counterpart at
wavelengths shorter than $\sim 0.8 \mu$m due to heavy interstellar
extinction. The system is persistently bright in soft and hard X-rays
(e.g. van der Klis 1993; Berger \& van der Klis 1994; Matz et al.
1996), with strong and variable metal emission lines (e.g. Liedahl \&
Paerels 1996; Kawashima \& Kitamoto 1996).  Several detections at
$\gamma$-ray energies have been claimed but rarely confirmed (see
e.g. Protheroe 1994).  A clear and persistent (observed for $> 20$ yr)
asymmetric modulation in the X-ray and infrared continuum emission
with a period of 4.8 hr (e.g. Mason, Cordova \& White 1986) is
interpreted as the orbital period of the system. This period is
rapidly lengthening with a characteristic timescale of less than a
million years (e.g. Kitamoto et al. 1995)

Infrared spectroscopy of the system in 1991 (van Kerkwijk et al. 1992)
first revealed the presence of broad emission lines and an absence of
hydrogen which was reminiscent of Wolf-Rayet stars. These observations
have subsequently been confirmed and expanded upon (van Kerkwijk 1993;
van Kerkwijk et al. 1996) and the binary interpreted as comprising a
compact object (neutron star or black hole) and the helium core of a
massive star, embedded within a dense stellar wind. Such an
evolutionary end-point was predicted for Cyg X-3 as far back as 1973
by van den Heuvel \& de Loore (1973).  Unfortunately most models of
Wolf-Rayet stars do not envisage objects which can be contained within
a 4.8 hr orbit, causing some dispute over this interpretation
(e.g. Schmutz 1993). Doppler-shifting of the broad emission lines with
the orbital period of the system, with maximum blue shift at X-ray
minimum, is interpreted by van Kerkwijk (1993) and van Kerkwijk et
al. (1996) as being due to the lines arising in the region of the
stellar wind shadowed from the X-rays of the compact object by the
companion star.  In this way the semi-amplitude of the Doppler-shifts
reflects only the wind velocity and gives no information on mass
function of the system.  Schmutz, Geballe \& Schild (1996) 
interpret the Doppler-shifting of the emission lines with the orbital
period more conventionally as tracking directly the motion of the
companion star and derive a mass function which implies the presence
of a black hole of mass $> 10 M_{\odot}$ in the system. However their
intepretation does not explain the phasing of the emission
lines relative to the X-rays, nor is this discrepancy addressed in
their work.

Mitra (1996, 1998) has argued that Cyg X-3 cannot contain
a massive W-R star as the optical depth to X-rays for a compact object
in a tight 4.8-hr orbit would be $>>1$. The alternative
explanation put forward is that Cyg X-3 instead contains a neutron
star and an extremely low-mass dwarf, cf. PSR 1957+20.

Van Kerkwijk (1993) discussed the dramatic variability in
line strengths and line ratios in the infrared spectra of Cyg X-3 and
suggested that when the source is bright in X-rays the emission lines
should be weak and orbitally modulated, but when the source is weak in
X-rays the lines should be strong and show little orbital
modulation. However, as noted in van Kerkwijk et al. (1996), Kitamoto
et al. (1994) show that the strength of infrared line and X-ray
emission are in fact probably broadly correlated from epoch to epoch,
with the strong lined spectrum of 1991 being obtained during an
outburst of the system. The explanation put forward for this was
enhanced mass loss from the companion during outbursts, which both
increases X-ray brightness (more accretion) and emission line
strengths. This model was combined with detailed radio, (sub)mm and
infrared (photometric) observations obtained during an outburst, and
expanded upon in Fender et al. (1997). Waltman et al. (1997) clearly
indicate the epochs of the published infrared spectra against the
Green Bank 2 GHz radio monitoring of the system.

In this paper we present four epochs of high-resolution infrared
spectroscopy of Cyg X-3 with the Multiple Mirror Telescope over a two
year period. These observations cover periods of quiescence, small
flaring and major outburst as revealed in radio and X-ray monitoring,
and we discuss the clear changes in the spectrum of the source as a
function of state.  In a future paper will analyze and discuss the
results of our spectra that fully sample the entire orbit of Cyg X-3
during quiescence and during outburst.

\begin{figure*}
\leavevmode\epsfig{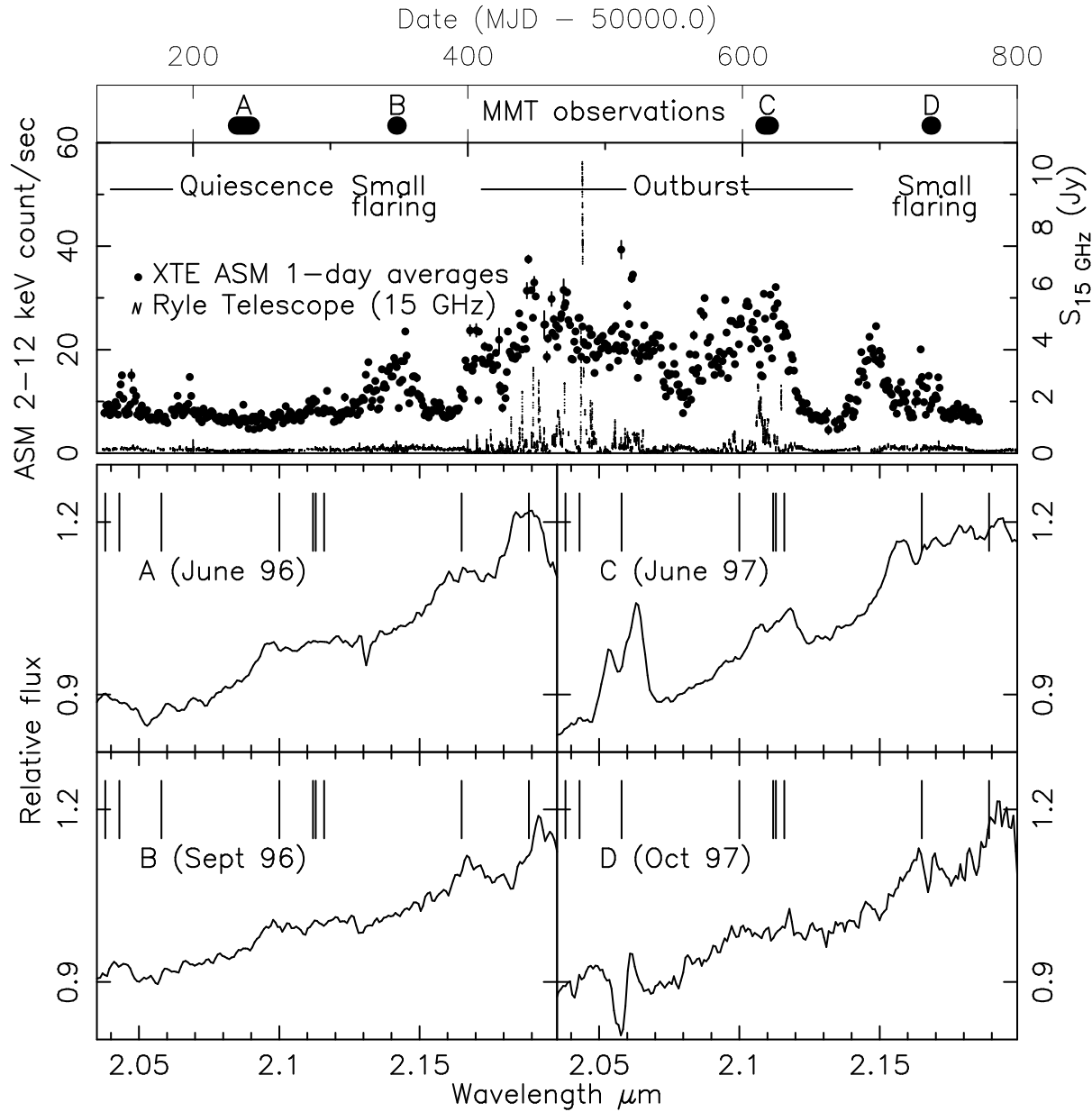}
\caption{An illustration of our four observing periods, labelled A --
D against a backdrop of X-ray and radio monitoring. We have observed
Cyg X-3 in distinct states of quiescence, small flaring and
outburst. Spectra characteristic of each epoch are indicated in the
lower panel, with tick marks indicating the lines identified in Fig 2.
The approximate S/N ratios for the spectra A--D are 80, 50, 100 and 40
respectively. Note that these spectra, unlike those presented in the
rest of the paper, have not been normalised to the continuum; this is
in order to show that there is no dramatic change in continuum slope
from outburst to quiescence.
}
\end{figure*}

\begin{figure*}
\leavevmode\epsfig{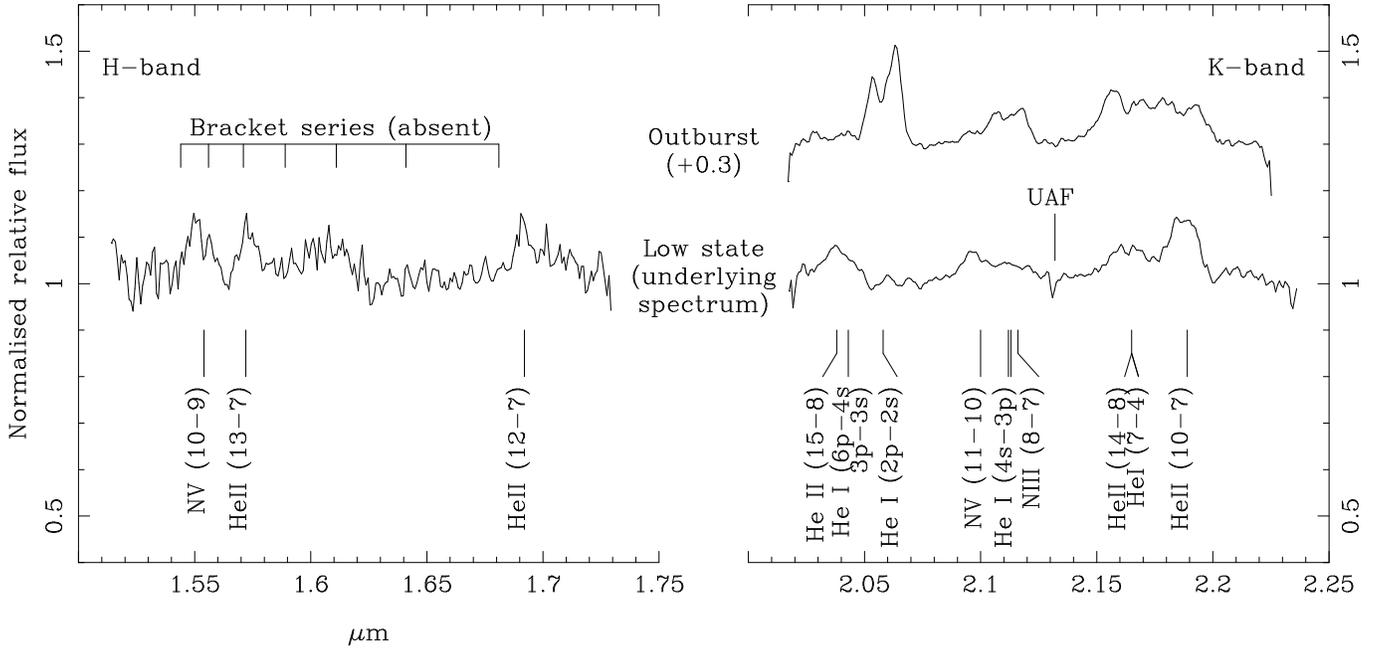}
\caption{Mean H-band (quiescent, taken 1996 June) and K-band
(outburst, upper taken 1997 June; quiescence, lower taken 1996 June)
spectra and line identifications.. UAF indicates a persistent (in 1996
June at least) unidentified absorption feature. The approximate S/N
ratios for the H-band, K-band (quiescent) and K-band (outburst)
spectra are 30, 100 and 80 respectively. These spectra, and those
presented throughout the rest of the paper, have been normalised to
the continuum.}

\end{figure*}

\section{Observations}

\subsection{Infrared}

All observations were made using the Steward Observatory's infrared
spectrometer, FSpec (Williams et al. 1993), on the Multiple Mirror
Telescope (MMT).  The spectra were taken using the medium resolution,
300 g/mm grating, yielding a 2 pixel resolution element of
0.0018~$\mu$m, or R $\approx$ 1200 at 2.12~$\mu$m, and R $\approx$ 900
at 1.62~$\mu$m.  The same observing procedure was used on all
nights. A full log of these observations is provided in Appendix A.

The spectrometer has a slit size of $1 \farcs 2 \times 32\arcsec$ on
the MMT, allowing Cyg X-3 to be observed in four unique positions
along the slit.  In the reductions, after dark current had been
subtracted and a flat field divided from the raw two-dimensional
images, sky emission and additional thermal background was removed by
subtracting one slit position from the next.  The integration times
for Cyg X-3 were very long, either 2 or 4 minutes at each slit
position (see tables in Appendix A). This is long enough that the
strong atmospheric OH emission lines did not always subtract away
cleanly due to temporal variations in atmospheric conditions between
slit positions.  In many cases, a few percent scaling was required to
get the OH features to disappear entirely.  Background normalization
of a few percent was performed to remove fluctuations in thermal
background between integrations.

Interspersed between our Cyg X-3 observations we obtained spectra
of other stars which were used to correct for telluric absorption 
features.  The same telluric standard star, HR 7826,
an A1 V, was used through out all observations.  The intrinsic 
spectrum of the standard star, HR 7826, was determined using two
secondary telluric standard stars, HR 7503 (16 Cyg A), a G1.5 V and
the O3 If*,  Cyg OB2 \#7.  A first estimate of the intrinsic 
spectrum of HR 7503 was obtained using a solar spectrum.  The 
2~$\mu$m spectrum of Cyg OB2 \#7 is nearly featureless, with the
exception of N~{\sc iii} at 2.115~$\mu$m (Hanson, Conti \& Rieke 1996).  
The A1 V telluric standard contains only the Br~$\gamma$ feature.  
By ratioing these three spectroscopically unique telluric standard 
stars against each other, we were able to obtain a good 
determination of the intrinsic spectrum of each star.  The intrinsic 
spectrum of HR 7826 was determined during our first observing run
in June 1996.  This solution for the intrinsic spectrum was used through 
out that run and with all future observing runs.  If our determination 
of the intrinsic spectrum of HR 7826 is not exactly correct, which
is certainly the case at some level, any spurious features 
we have introduced will at least be {\sl consistently} introduced
into all of our Cyg X-3 spectra.  This is important since it is our
hope to study flux and velocity variations of very weak broad features
in Cyg X-3 in an upcoming paper.  

Mean spectra for the four epochs of observation, and their relation to
the changing X-ray and radio state of Cyg X-3, are shown in Fig 1.
Note that these spectra are not normalised, whereas those throughout
the rest of the paper are. This is in order to show the approximate
constancy of the continuum slope in different states. The correlation
between radio flaring and bright X-ray states, originally proposed by
Watanabe et al. (1994) is also obvious from Fig 1.

We began our first Cyg X-3 observing campaign in late May, 1996, which
is symbolized in Figure 1 as epoch A.  For ten of the eleven
consecutive nights, Cyg X-3 was observed at approximately the same UT.
Because a 24 hour daily cycle is almost exactly five binary orbits, we were
observing Cyg X-3 at close to the same orbital phase for these ten
nights (see Table 1 in Appendix A).  Furthermore, on 2 June 1996, Cyg
X-3 was observed over an entire orbital period, from $\phi_X$ = 0.185
to 1.181 (quadratic ephemeris of Kitamoto et al.\ 1995, where $\phi_X
= 0$ corresponds to minimum X-ray flux in the 4.8 hr modulation,
probably the point of superior conjunction of the compact object).
During this first campaign, $H$-band spectra centred at 1.62~$\mu$m
were also obtained on the 7th and 8th of June 1996 (see Fig 2).

The second campaign of observations, represented by epoch B in Figure 1,
began 22 September 1996, where we obtained spectra covering the entire
orbital period, from $\phi_X$ = 0.382 to 1.429.  One fifth of an orbit
was observed the following night (Table 2 in the Appendix).  The third 
observing campaign covered five consecutive nights beginning 16 
July 1997 (Table 3 in the Appendix) and are represented 
in Figure 1 as epoch C.  The fourth night of the observations 
taken during epoch C covered one orbit, sampling from $\phi_X$ 
= 0.205 to 1.109.   Our final observing campaign, represented by epoch
D in Figure 1, covered just one quarter of an orbital period on 15 
October 1997.

\subsection{Radio}

The Ryle Telescope observations, at 15 GHz with a bandwith of 350 MHz,
follow the pattern described in Fender et al (1997).  Data points
shown in Figs 1 and 3 are 5-min integrations. The typical uncertainty
in the flux-density scale from day to day is 3\%, and the rms noise on
a single integration is less than 2 mJy.

\subsection{XTE}

Cyg X-3 is monitored up to several times daily in the 2-12 keV band by
the Rossi XTE All-Sky Monitor (ASM). See e.g. Levine et al. (1996) for
more details. The total source intensity in the 2-12 keV band for
individual scans is plotted in the top panels of Figs 1 and 3.

\section{Line Identifications}

Line identifications in Cyg X-3 are shown in Figure 2 and listed in
Table 1.  We display two different $K$-band spectra in Figure 2, the
upper taken during a time of high x-ray and radio activity, the lower
taken during quiescence.  The strongest features include the
2.0587~$\mu$m He~{\sc i} singlet during outburst and the 2.1891 He~{\sc ii} (7-4)
during quiescence.  The $H$-band spectrum centered at 1.62~$\mu$m,
displays only a few identifiable features, He II (13-7) and (12-7) at
1.5719 and 1.6931$\mu$m, respectively, and N~{\sc v} (10-9) at 1.554~$\mu$m.
These H-band features were also evident in earlier UKIRT spectra 
from 1992 May 30, one day after K-band spectra revealed Cyg X-3 to be in
a weak-lined state equivalent to quiescence as defined in this paper 
(M.H. van Kerkwijk private communication).
There is no evidence for any Brackett series hydrogen features.  The
$H$-band spectrum shown in Figure 2 was taken in June 1996, when Cyg
X-3 was in a quiescent phase.

\begin{table*} 
\begin{minipage}{180mm}
\centering
\caption{Line Identifications and Equivalent Widths}
\begin{tabular}{lllcccc}  
\hline 
Ion     & Transition  &       Vacuum       &
\multicolumn{2}{c}{Equiv. Width (\AA)} & \multicolumn{2}{c}{FWHM (\AA
/ km s$^{-1}$)} \\
        &             & $\lambda$ ($\mu$m) & Quiescence$^1$ &
Outburst$^1$ & Quiescence$^1$ & Outburst$^1$ \\
\hline
N~{\sc v}      & (10-9)      &      1.554         &     -15  &     --&
130/1500 & --  \\
He~{\sc ii}    & (13-7)      &      1.5719        &     -12  &     --&90/1700 & -- \\
He~{\sc ii}    & (12-7)      &      1.6931        &     -9   &     --&80/1400
& --\\
He~{\sc ii}    & (15-8)      &      2.0379        &    -10   &    $>$1 &
150/2200 & 70/1000 ? \\
He~{\sc i}     & 6$p\, ^3P-4s\, ^3S$ & 2.0430$^2$ &  blended &    -1? & blended & ? \\
He~{\sc i}     & $2p\, ^1 P-2s\, ^1S$  & 2.0587 &    $>$2  &    -25 & ? & 150/2200\\
N~{\sc v}      & (11-10)     &      2.1000        &     -7   &    -3 &
120/1700 & 150/2200 \\
\multicolumn{3}{l}{$\left. \begin{array}{lll} 
{\rm He~{\sc i}}  \,  & 4s\, ^3S-3p\, ^3P & \ \  2.1126^2  \\
{\rm He~{\sc i}}  \,  & 4s\, ^1S-3p\, ^1P & \ \ 2.1132  \\
{\rm N~{\sc iii}}    & (8-7)       &   \  \,  2.1155  \end{array} \right\}$ }
& -2   & -12 & blended & blended\\   
UAF     &             &      2.129         &      +1      &   $>$1 &
17/$<$250 & 18/$<$250 \\
\multicolumn{3}{l}{$\left. \begin{array}{lll}
{\rm He~{\sc ii}} \,  & (14-8)   \ \ \ \ \   &   2.1653       \\        
{\rm He~{\sc i}}  \   & (7-4)$\phantom{00'0000}$ &  2.1655^2 \end{array} \right\}$ }
& -7 & -25 & blended & blended \\
He~{\sc ii}    & (10-7)      &      2.1891        &   -15         &    -12 &
140/1900 & 200/2700\\
\hline
\multicolumn{5}{l}{$^1$Low state observed June 1996, Outburst observed June 1997.} \\
\multicolumn{5}{l}{$^2$Blended; line centre not well enough
constrained to determine precise electronic transitions} 
\\ 
   
\end{tabular}
\end{minipage}
\end{table*}

There is one absorption feature, centered at approximately
2.129~$\mu$m, that we have been unable to positively identify.  It is
unlikely that it is a feature due to intervening interstellar
material, as numerous stars with line of sight extinction greater than
ten magnitudes in the visible have been observed without ever showing
such a feature (Tamblyn et al.\ 1995; Hanson, Howarth \& Conti 1997;
Watson \& Hanson 1997).  We suspect then, it must be related
to the Cyg X-3 system.  Curiously, it shows no shifting with the
orbit, unlike the other lines in the K-band (with the possible
exception of He~{\sc i} at 2.058~$\mu$m).  This unidentified
absorption feature (UAF), has since disappeared from the spectrum,
starting in June 1997.  We have seriously considered that the feature
may be spurious, introduced by poor telluric corrections, or perhaps a
bad pixel on the array.  However, we see it present through out the
entire 11 day run in 1996 June, despite small changes in grating
position, against three different telluric standard stars, and new
calibration images taken each day.  Furthermore, inspection of earlier
2~$\mu$m spectra of Cyg X-3, while of lower resolution, seems to
substantiate the presence of a weak absorption feature at 2.129~$\mu$m
(van Kerkwijk et al. 1996).  However without an identification, we are
unable to comment further on its nature or its possible relation to
the Cyg X-3 system.

\section{Spectral variability}

In this section we discuss the observational properties at each of the
four epochs for when near-infrared spectra were obtained.  It is our
aim to establish the spectral characteristics and nature of any variations
seen in Cyg X-3 in different radio and X-ray states.  This may help us 
to identify the origin of the spectral features, be they from the secondary
star or the compact object. Spectra characteristic of each epoch are 
plotted in Fig 1.

\subsection{1996 May / June : quiescence}

Represented by epoch A in Figure 1, this is the longest continuous set
of near-infrared observations ever taken of Cyg X-3. The source is in
a state of radio and X-ray quiescence, with radio flux densities at 15
GHz in the range 40 -- 140 mJy and XTE ASM fluxes in the range 4 -- 9
count/sec. The spectrum is dominated by broad weak He~{\sc ii} and N~{\sc v}
emission, and weak, more narrow and intermittent He~{\sc i} (2.058 $\mu$m)
absorption. Nearly all spectral variability is related to the 4.8 hr
orbital modulation, namely Doppler-shifting of the broad emission
features.  The full amplitude of the Doppler-shifting is of the same
order as that reported by van Kerkwijk (1993) and Schmutz et
al. (1996), i.e. $1000$ --  $1500$ km s$^{-1}$.  Orbitally phase-resolved
spectra and dynamical interpretations will be presented elsewhere. The
unidentified absorption feature (UAF) at 2.129 $\mu$m is also
detected, but cannot be clearly identified with any known transition.
The UAF shows no Doppler-shifting.

\subsection{1996 September : small flaring}

The second set of observation, epoch B in Figure 1, caught Cyg X-3 in 
a more active phase. Radio observations at 15 GHz showed many small
flares with flux densities ranging from 50 -- 450 mJy, corresponding
to the `small flaring' state classified by Waltman et al. (1995).
The XTE ASM recorded 8 -- 17 count/sec, significantly higher and more
variable than in 1996 May / June. However, the K-band spectrum is very
similar to that obtained at epoch A, showing little variability that
is not orbitally-related, and being dominated by the broad weak He~{\sc ii}
and N~{\sc v} emission. The unidentified absorption feature at 2.129 $\mu$m
appears to have weakened considerably in the three months since 1996 
May / June.

\begin{figure}
\leavevmode\epsfig{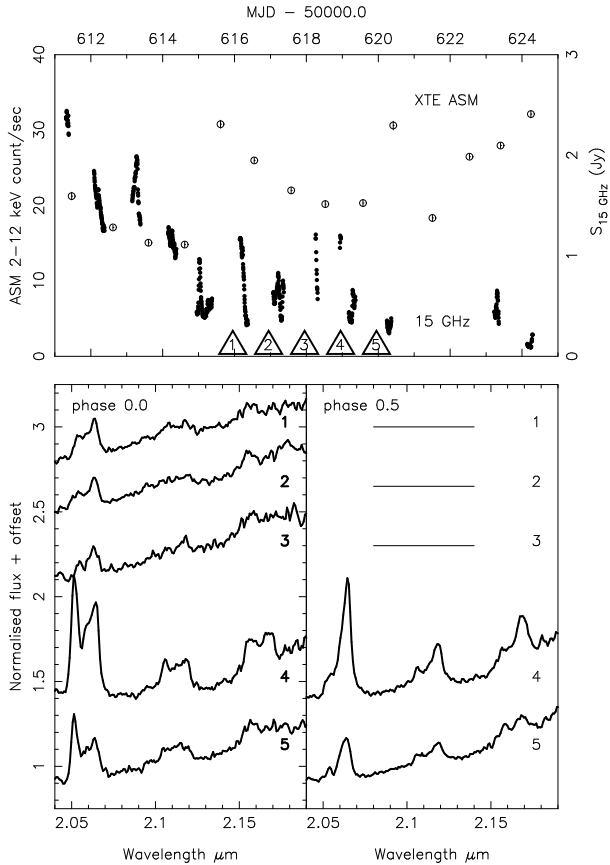}
\caption{Spectra on five consecutive nights during outburst in 1997
June. Top panel shows XTE ASM 2-12 keV (open symbols) and 15 GHz radio
monitoring (filled symbols), revealing the source to be in a bright
and variable state, presumably arising from enhanced accretion and jet
formation. All spectra are dominated by apparently twin-peaked He~{\sc i}
emission.  Day 1 corresponds to 1997 June 16.  Days 4 \& 5 reveal the
development and subsequent decline of especially strong emission. This
structure displays the same blue : red wing variability with orbital
phase on two subsequent days, being twin-peaked around phase zero but
very red-dominated half an orbit later.}
\end{figure}

\subsection{1997 June : outburst}

Epoch C represents observations during a major outburst of Cyg
X-3. XTE ASM count rates varied rapidly between 14 -- 32 ct/sec, having
peaked at $\geq 40$ ct/sec around 100 days earlier.  The radio
emission was undergoing a second sequence of major flaring within 200
days. During the period of these observations flux densities of up to
3 Jy at 15 GHz were recorded. During the first period of radio flaring
(MJD 50400 -- 50500) Mioduszewski et al. (1998) clearly resolved an
asymmetric, probably relativistic, jet from the source.

The K-band spectrum at this epoch is wildly different from that at any
other epoch, being dominated by what appear to be very strong
double-peaked He~{\sc i} emission features, most obviously at 2.058 $\mu$m.
Significant day-to-day spectral changes which are not related to
orbital phase are evident at this epoch, unlike in quiescence (where
spectral variability is almost entirely due to orbital modulation --
see above).

Figure 3 illustrates the dramatic variability in the strength of the
double-peaked He~{\sc i} emission over the five nights of observations :
strongest emission is present on the fourth night, 1997 June 19.  Fig
4 shows in detail the rapid V/R variability, probably cyclic at the
4.8 hr orbital period, observed on this date.  Figure 3 also hints at
a possible anti-correlation between 2-12 keV X-ray flux and He~{\sc i}
emission line strength on day timescales; while there is a large
degree of X-ray variability in individual scans, the daily-averaged
flux drops during the first four days to a minimum just after June
19. This is in contrast to the longer-term correlation between X-ray
state and He~{\sc i} line strength, and probably results from a drop in
ionisation state of the wind following a temporary decrease in X-ray flux.

\begin{figure}
\leavevmode\epsfig{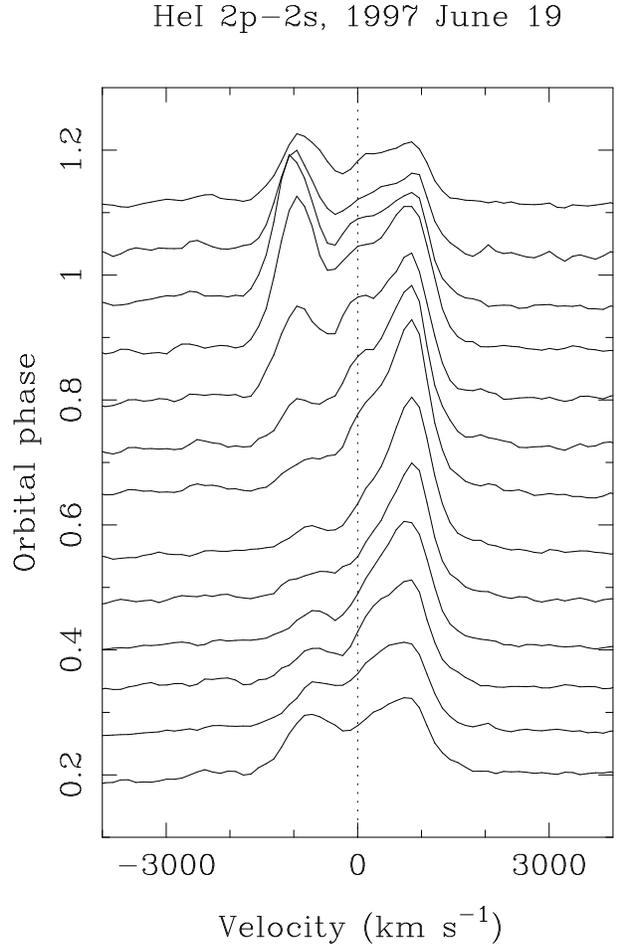}
\caption{Rapid V/R variability in the He~{\sc i} 2p-2s (2.0857 $\mu$m)
emission line on 1997 June 19 (day 4 in Fig 3). The variability is
probably cyclic at the 4.8 hr orbital period. The red-shifted peak is
far more persistent than the blue-shifted peak, suggestive of some
continuous moderately blue-shifted absorption.}
\end{figure}

The dramatic day-to-day variability illustrated in Figure 3 suggests
that the lines seen are transient features which are not tied to a
steady-state wind of the secondary. The wind is most likely changing
in ionisation state and/or density/velocity on day timescales.  The
strongest peak seen in He~{\sc i} on day 4, showing up blue-shifted at
$\phi_X$ = 0.0 and red-shifted at $\phi_X$ = 0.5, is clearly evolving
on a timescale of days, more than tripling in strength between days 3
-- 4, and declining again within 24 hr.  The appearance of the
1.0830~$\mu$m He~{\sc i} feature on 14 June 1993 (van Kerkwijk et al.\ 1996)
likely represented another one of these events, though 2 ~$\mu$m
spectra are not available to confirm.  van Kerkwijk et al.\ (1996) do,
however, show there was a marked increase in K-band flux on 14 June
1993.  Such near-infrared flux increases on day timescales have
been seen during radio outbursts (Fender et al.\ 1997), and appear to
be distinct from the more rapid (second to minutes timescales)
infrared flaring which is often observed (e.g. Fender et al. 1996).

\subsection{1997 October : post-outburst / small flaring}

A single short (64 min total) observation on 1997 October 15 ($\phi_X$
= 0.58 - 0.81), during an apparent decline to quiescence following
outburst, epoch D again reveals previously unobserved
features. Alongside the quiescent weak broad He~{\sc ii} and N~{\sc v}
emission is strong He~{\sc i} 2.0587 $\mu$m absorption, displaying a
P-Cygni profile. This absorption is stronger than observed at any time
during epoch A. The absorption is present in all individual spectra,
and there is no evidence for significant variability on short
(minutes) time scales. The absorption minimum occurs within
uncertainties at the rest wavelength of the transition, 2.058 $\mu$m,
and the blue wing extends to $\sim$ 2.054 $\mu$m, implying a minimum
outflow velocity of 500 km s$^{-1}$. The He I absorption feature does
not display any Doppler-shifting, though our phase coverage is not
ideal.  There are no other significant absorption features in the
spectrum.  The UAF feature is entirely absent.

A comparison of the spectrum around the He I 2.0587 $\mu$m with that
in outburst (Fig 5) shows that the deep absorption may well be present
in outburst also, but is completely dominated by much enhanced
emission at this stage. This is compatible with the model for a
disc-like wind which we explore in section 6 below.

\section{Discussion}

\subsection{The Near-Infrared Spectral Type of the Secondary}

Van Kerkwijk et al.\ (1992) published the first near-infrared spectra
of Cyg X-3, covering 0.72-1.05~$\mu$m (I-band) and 2.0-2.4~$\mu$m
(K-band).  These spectra, taken in late June 1991, displayed strong
emission lines of He~{\sc i} and He~{\sc ii}.  The I-band
spectrum in particular, showed a conspicuous absence of hydrogen
lines.  The lack (or much reduced fraction) of hydrogen, the strong
He~{\sc i} emission at 2.058~$\mu$m, and the broad He~{\sc ii} emission 
lines were interpreted as coming from the wind of the binary
companion to the compact object in Cyg X-3.  Based on the 1991
spectrum, and using comparison spectra obtained of several Wolf-Rayet
stars which were observed at the same time, a spectral type of WN7 was
estimated for the companion.  There are some problems, however, with
the June 1991 spectra.  The spectrum showed strong narrow He~{\sc i} 
at 2.0578~$\mu$m with strong, broad He~{\sc ii}, which is not generally
seen in hydrogen-free WN Wolf-Rayet stars (Figer, McLean \& Najarro 
1997; c.f. WR 123 in Crowther \& Smith 1996).  
This subtle mis-match of spectral characteristics
suggested that the lines seen in the original June 1991 spectrum did
not originate solely from a WR-like wind. Indeed, subsequent spectra
taken by van Kerkwijk et al. (1993) showed that the originally strong
He~{\sc i} features had since disappeared.  These later spectra were
now dominated by the broad He~{\sc ii} features, as well as N~{\sc
v} and N~{\sc iii}.  Such features are indicative of an earlier WR wind,
perhaps WN4/5.  However, as noted by van Kerkwijk et al.\ (1996), the
line ratios between the nitrogen and He~{\sc ii} lines are not
consistent with such an early spectral class.  In fact, the
near-infrared He~{\sc ii} lines in Cyg X-3 are extremely weak compared
to other early WN stars (Crowther \& Smith 1996; Figer et al.\ 1997).

We are now able to show that the original 1991 June spectrum of van
Kerkwijk et al. (1992) was anomalous and almost certainly associated
with an outburst in the system.  Our 1997 June spectra are dominated by
double peaked emission, which does not seem to be traced in the
original 1991 spectrum.  However, by choosing a phase that was
dominated by one peak and smoothing our spectra to the lower
resolution of the van Kerkwijk et al. (1992) spectrum, our 1997 June
spectra became a very close match in both lines detected and
relative strength to the 1991 spectrum (Fig 6).  As already suspected
by van Kerkwijk et al. (1996), the original K-band spectrum of van
Kerkwijk et al. (1992) therefore appears to have been anomalous due to
an outburst of Cyg X-3.

The quiescent spectrum, dominated by weak, broad He~{\sc ii} features,
likely originates in the more steady-state wind of the stellar
companion of Cyg X-3 and is our best diagnostic of the nature of this
component.  However, even this phase is not consistent with a normal
WR wind.  As first suggested by van Kerkwijk (1993), the presence of
the high energy compact object, circling the companion star at very
close radii (estimated to be on the order of 5 R$_{\odot}$), has
likely altered the wind structure of its companion (the ``Hatchett
McCray effect,'' Hatchett \& McCray 1977). Stellar winds in early-type
stars are driven through high opacity resonance lines of such species
as C~{\sc iv} and N~{\sc v}.  The predominance of very high energy
photons from the compact object completely alters the ionization
structure and thus the driving force of the wind, and may entirely
eliminate significant line formation (McCray \& Hatchett 1975).  In
the presence of the compact object, an X-ray-excited, thermally driven
wind is instead created, which may have little or no line formation
(Stevens 1991; Blondin 1994).  Where the compact object is entirely
blocked by the central disk of the companion, the expanding wind from
the helium star may be capable of creating a normal line-driven wind,
giving rise, at least weakly, to the broad high ionization wind lines
detected in Cyg X-3.

With such an interpretation for the line emission seen at
near-infrared wavelengths, it would be difficult to infer many
characteristics of the companion star.  The most important
characteristics of the companion star are that it is a helium rich
atmosphere, and it may be driving a fairly extensive, fast wind, both
being reminiscent of late-stages in massive star evolution.
An early WN Wolf-Rayet star is likely the best candidate for the
spectral type of the companion star. However,
the mass of the companion star, and thus information on the mass of
the compact object, can not be uniquely or confidently determined from the
spectrum. 

\subsection{Twin-peaked He I emission in outburst}

Here we discuss possible origins for the strong twin-peaked He~{\sc i}
emission observed during outburst.

\begin{figure}
\leavevmode\epsfig{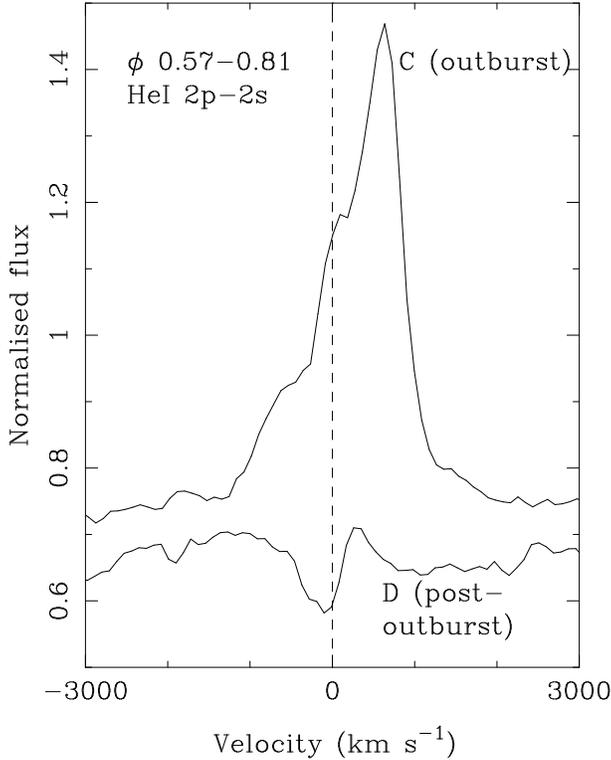}
\caption{ A comparison of outburst (epoch C, 1997 June 19) and
post-outburst (epoch D, 1997 Oct 15) spectra around the He~{\sc i}
2p-2s line at 2.0587 $\mu$m, summed in the phase interval
0.57-0.81. It seems plausible that the strong P-Cyg absorption to -500
km s$^{-1}$, clearly evident in the post-outburst spectrum, is present
at both epochs, and the only difference may be vastly reduced amount
of emission in 1997 October.  }
\end{figure}

\begin{figure}
\leavevmode\epsfig{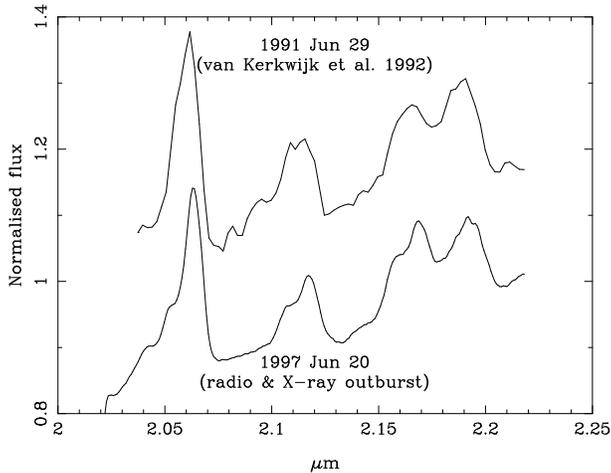}
\caption{A comparison of the 1991 Jun 29 spectrum of van Kerkwijk et
al. (1992) with our spectrum of 1997 June 20, obtained during a radio
and X-ray outburst.  The similarity of the two spectra (contrast with
the quiescent spectra in Figs 1 and 2) indicates that the initial WN7
classification based upon van Kerkwijk's spectrum was not
representative of the underlying spectral type of the companion, but
instead a result of enhanced He~{\sc i} emission during outburst.}
\end{figure}

\subsubsection{Jets ?}

Simple arguments show that the twin-peaked emission lines cannot arise
directly from material in the relativistic outflows which are inferred
from high-resolution radio mapping of Cyg X-3 (e.g. Geldzahler et
al. 1983; Mioduszewski et al. 1998). Firstly, the persistently
stronger red wing as shown in Figure 5 is the opposite of what would
be predicted from Doppler boosting, where the approaching (blue
shifted) emission would be boosted, and the receding component
diminished.  Secondly, the relatively low velocity ($\sim 1500$ km
s$^{-1}$) implied by the peak separation could only arise from a
relativistic jet almost in the plane of the sky (i.e. with a very
small radial component). In that case, the transverse Doppler shift
due to time dilation would dominate, red shifting both components.
For a velocity of around 0.3 c this would result in a red shift of
both components by $\sim 0.1$ $\mu$m, which is not observed (note this
effect {\em is} observed in SS 433 - see e.g. Margon 1984). The lack
of a discernible transverse red shift (assuming the lines do
correspond to the indicated He I transitions) effectively rules out a
relativistic outflow.

A lower-velocity non-relativistic jet is possible, although it still
suffers from the problem of explaining the persistently stronger red
peak, but we consider this unlikely as rapidly variable radio
emission is occurring throughout this period (Fig 3). This is almost
certainly associated with the production of a relativistic jet; given
the existence of this jet and the strong wind the presence of a third
outflowing component (which matches the terminal velocity of the wind
as inferred from quiescent observations) seems unlikely.

\subsubsection{Accretion disc ?}
    
As already noted by van Kerkwijk et al. (1992), a significant
contribution to the infrared emission of Cyg X-3 from an accretion
disc is unlikely. This is because in order to generate the observed
infrared luminosity the disc would need to be very hot ($\geq 10^{6}$
K), as its size is tightly constrained by the 4.8 hr orbit. Such a
high temperature is hard to reconcile with the observed low-excitation
He~{\sc i} features.  However, the line profiles and velocity separation are
reminiscent of features seen in optical spectra of accretion-disc
dominated systems, and it is worth checking in more detail.

We can calculate the temperature that a black body (the most efficient
emitter) would require in order to reproduce the observed line flux,
given that its size is constrained by the dimensions of the orbit.  We
assume a distance of 8.5 kpc, a binary separation of $5 R_{\odot}$, a
K-band extinction $A_K = 2.3$ mag, and a flux in emission lines which
is about 10\% of that in the continuum. For an observed flux density
in the K-band of $\sim 12$ mJy (e.g. Fender et al. 1996) we find that
we require a black-body temperature in excess of $10^6$ K in order to
produce the flux in the emission lines within the binary
separation. As the emission of the plasma producing the lines is much
less efficient than that of a black body, there seems to be no way in
which the relatively low excitation He~{\sc i} lines can be produced
within the scale of the binary separation, as these lines need
temperatures $T \leq 10^5$ K (for reasonable densities). Using this
temperature we can find a minimum dimension for the emitting
region. As $r \propto T^{-1/2}$ we require an emitting region which is
a factor of three larger, i.e. $\sim 15 R_{\odot} = 10^{12}$cm. Such a
large separation for a 4.8 hr orbit would imply a total mass in the
system of 1000 $M_{\odot}$ ! So, we can rule out an emitting zone
which is contained within the orbit of the system.

Furthermore, the luminosity of the emission lines, both in outburst
and quiescence, is orders of magnitude greater than that observed in
K-band emission lines from the X-ray binary Sco X-1 (Bandyopadhyay et
al. 1997). Given that Sco X-1, with a longer orbital period, probably
possesses a larger (and hence brighter in the infrared) accretion disc
than Cyg X-3, an origin for the infrared lines of Cyg X-3 in an
accretion disc can be ruled out (unless the distance is overestimated
by a factor of 10 or more - which seems highly unlikely given the
broad agreement between high optical/infrared extinction, high $N_H$
in X-ray spectral fits, and the distance inferred from 21 cm radio
observations).

So, in agreement with van Kerkwijk et al. we must conclude that {\em
the He~{\sc i} emission lines arise from a region significantly larger than
the binary separation of the system}. This conclusion also rules out
an origin for the emission lines in the X-ray irradiated face of a
relatively cool secondary.

\subsubsection{An enhanced, possibly disc-like wind ?}

Here we discuss a third possible origin for the twin-peaked variable
emission lines : a significant enhancement in the wind in Cyg X-3.
This has already been suggested as the origin for outbursts from the
system (Kitamoto et al. 1994; van Kerkwijk et al.\ 1996; Fender et
al. 1997).  Given that we have established that the twin-peaked
emission lines almost certainly originate in an extended region which
is not the jets, and the existing evidence for a strong wind in the
Cyg X-3 system, a natural explanation is that the increased line
strength in outburst represents an increase in the density of the
WR-like wind in the system.  Such an increase in density will be
coupled to a decrease in the mean ionisation level of the wind, hence
the much increased He~{\sc i} : He~{\sc ii} ratio. While an enhanced
wind density of the companion star is a natural explanation for bright
X-ray / radio states which reflect increased rates of accretion and
jet formation, such enhancements have never been observed in other WN
stars. Cyg X-3, however, is an exceptional system. It experiences both
extreme tidal forces and irradiation, which likely induce erratic
behavior and non-periodic variations in the extended atmosphere of the
companion Helium star.

\begin{figure*}
\leavevmode\epsfig{file=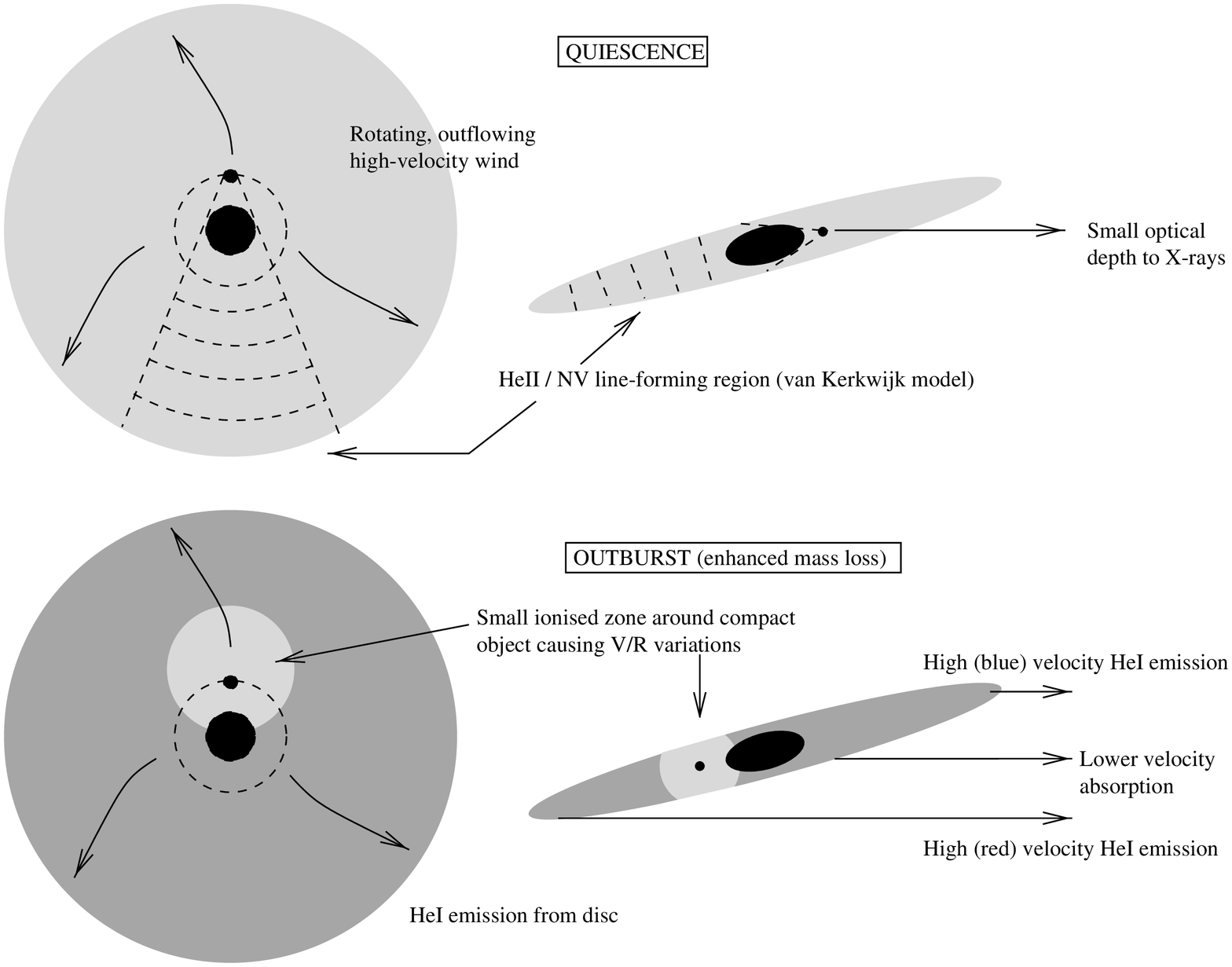,angle=270,width=17cm,clip}
\caption{A disc-wind in the Cyg X-3 system. 
In quiescence orbital
modulation follows essentially the model of van Kerkwijk (1993) and
van Kerkwijk et al. (1996), with the X-ray source ionising the entire
wind except that region shadowed by the companion.
In outburst, caused by enhanced mass-loss
from the companion star, the X-ray source can only ionise a small
local region (Stromgren zone) and He~{\sc i} emission dominates. V/R
variability and asymmetry is caused by P-Cygni absorption from the
accelerating region of the wind seen against the companion, and by the
Stromgren zone as it tracks the X-ray source around the 4.8-hr
orbit. This disc-wind model can reconcile a massive, Wolf-Rayet-like
companion with a small optical depth to X-rays.}
\end{figure*}

The appearance of the twin-peaked lines, and their
variability (probably) in phase with the 4.8 hr orbit suggests an
origin in an asymmetric emitting region. We believe that this wind may
be flattened and disc-like, probably in the plane of the binary (see
e.g. Stee \& de Ara\'{u}jo 1994 for predicted line profiles from a
disc-wind). A flattened wind may have formed in the Cyg X-3 system as a
result of a rapidly (synchronously) rotating mass donor and/or
focussing of non-accreted material into the binary plane by the
compact object. In this case most of the infrared emission arises from
material in the plane of the binary but {\em outside} the orbit, and
the optical depth along the line of sight from the X-ray source to the
observer remains small as long as the system is not viewed edge-on
(see Fig 7).  In this way, the problem of reconciling the Wolf-Rayet
spectral typing of the companion with the detection of X-ray emission
from near the centre of the system, highlighted by Mitra (1996, 1998),
can be side-stepped whilst also explaining the large infrared
luminosity of the system. It is worth recalling however that Berger \&
van der Klis (1994) show from timing studies that the X-ray emission
from Cyg X-3 must still be undergoing significant scattering. 

Further support for a disc-wind model may come from the infrared
polarimetric observations of Jones et al (1994) who found a
significant degree of intrinsic polarisation from Cyg X-3 in the
infrared K-band. They suggested that this may indicate a preferential
plane of scattering in the binary. Several WR stars also show
intrinsic polarisation, interpreted as arising from scattering in a
flattened wind (e.g. Schulte-Ladbeck, Meade \& Hillier 1992;
Schulte-Ladbeck 1995; Harries, Hillier \& Howarth 1998). Such
intrinsic polarisation seems to be more common from WN type
Wolf-Rayets (Schulte-Ladbeck et al. 1992; Harries et al. 1998); and
the only direct observation (radio interferometry) of a flattened
Wolf-Rayet wind was also from a WN subtype (Williams et al. 1997).
Additionally, the position angle of the radio jet in Cyg X-3
(e.g. Mioduszewski et al. 1998) is approximately perpendicular to the
long axis of the flattened wind as inferred from the position angle (0
-- 40 degrees) of the derived intrinsic infrared
polarisation. Assuming the jet propagates along the axis of the
accretion disc, which itself lies in the binary plane, this supports a
model in which binary and wind planes are aligned.

We discuss the interpretation of the outburst state in the context of
a flattened disc-like wind in more detail below.

\section{Orbital modulations with a disc-wind}

\subsection{Quiescence}

Our disc-wind model for Cyg X-3 is sketched in Fig 7.  In such a
model, Doppler-shifting of He~{\sc ii} and N~{\sc v} lines in
quiescence would occur essentially as outlined in the model of van
Kerkwijk (1993) and van Kerkwijk et al. (1996) (hereafter the `van
Kerkwijk' model). We expect most of this emission to arise from
outside the binary, with the compact object orbiting within the
wind-accelerating zone of the WR-like companion. As discussed in the
Introduction and in van Kerwijk et al. (1996) and Mitra (1998) the van
Kerkwijk model naturally explains the phasing of the X-ray and
infrared continuum modulation with the Doppler shifts, in contrast to
the model of Schmutz et al. (1996).

\subsection{Outburst}

During outburst, we presume that the much-enhanced mass loss and
consequent higher wind density prevents the X-ray source from ionising
anything but a small fraction of the wind (unlike in the van Kerkwijk
model for quiescence in which the majority of the wind is ionised).  A
quantitative level of enhancement above the quiescent state is
difficult to estimate.  A realistic model to describe the increased
He~{\sc i} emission would require knowledge of the geometry and
structure (the clumpiness) of this wind, as well as the fractional
increase in mass loss rate.  An increase in the soft X-ray flux by a
factor of three during outburst indicates a corresponding increase in
the mass accretion rate during such periods, although density
enhancements close to the compact object may not exactly reflect those
in the wind as a whole. It might be possible, given arguments based on
the time scale of the line structures seen, to estimate a fractional
density of the wind during outburst.  Such an in depth analysis is
beyond the scope of this study. Possibly the most accurate measure of
the degree of wind enhancement may come from measurements of
`glitches' in the orbital period derivative as angular momentum is lost
from the system at a higher rate during the outbursts.
 
The rapid, probably cyclic V/R variability observed during outburst
could occur as a combination of three components

\begin{enumerate}
\item{Broad He~{\sc i} emission from the entire disc-wind, from approximately
-1500 to +1500 km s$^{-1}$.}
\item{An ionised region (Stromgren zone) local to the X-ray source
which depletes the He~{\sc i} emission in that region of the orbit.}
\item{Lower-velocity ($\sim -500$ km s$^{-1}$) blue-shifted (P-Cygni)
absorption from the accelerating region of the wind observed against
the companion star.}
\end{enumerate}

This simple scheme (illustrated in the lower panel of Fig 7) can
qualitatively explain the observed phasing of the V/R variability and 
the greater persistence of the red-shifted peak :

\begin{itemize}
\item{{\bf Phase 0.0 :} X-ray source is on far side of wind from the
observer. Red-shifted peak is depleted at relatively low velocities
due to ionised zone around X-ray source. Similarly blue-shifted peak
is depleted at lower-velocities due to persistent P-Cyg absorption.}
\item{{\bf Phase 0.5 :} X-ray source is on near side of
wind. Blue-shifted emission is depleted both by P-Cyg absorption and
ionisation from X-ray source; red-shifted peak is unaffected by either
and is much stronger.}
\end{itemize}

In the context of this model, the spectrum obtained in 1997 Oct (epoch
D, Fig 5) still shows deep P-Cyg absorption but much-reduced emission. This
may represent an intermediate state in the return to quiescence in
which the the low-velocity absorption is still occuring in the densest
parts of the wind, but beyond the binary orbit most of the material is
ionised and He~{\sc ii} / N~{\sc v} dominate over He~{\sc i} as in quiescence.

\section{Conclusions}

We have presented the most comprehensive and highest-resolution set of
infrared spectra of Cyg X-3 to date. In combination with X-ray and
radio monitoring we can characterize the infrared spectral behaviour
of the source in outburst and quiescence.

The underlying infrared spectrum of Cyg X-3, observed during both
radio and X-ray outburst and quiescence, displays weak, broad, He~{\sc
ii} and N~{\sc v} (but no He~{\sc i}) emission. Some He~{\sc i} 2.058
$\mu$m absorption may be present, preferentially around orbital phase
zero. H-band spectra extend our spectral coverage and confirm the
significant He-enrichment of the mass donor, with no evidence of any
hydrogen features. While not perfect, the closest match to the
spectrum is that of a hydrogen depleted early WN-type Wolf-Rayet star.

In outburst, the K-band spectrum becomes dominated by twin-peaked He~{\sc i}
emission, which is shown to be unlikely to arise in relativistic jets
or an accretion disc. This emission seems to arise in an enhanced wind
density, presumably also responsible for the X-ray and radio outburst
via enhanced accretion and related jet formation. This explains the
observed long-term (outburst timescale) correlation between emission
line strength and X-ray and radio state, as noted in Kitamoto et
al. (1994).  The emitting region almost certainly extends beyond the
binary orbit, and displays significant day-to-day intensity
variations, as well as V / R variability with orbital phase. The short
term (day-to-day) variability in He~{\sc i} line strength may be
anticorrelated with X-ray flux due to a varying degree of ionisation
of the wind. It seems that, for Cyg X-3 at least, the major X-ray and
radio outbursts are due to mass-transfer, and not disc, instabilities.
If this interpretation is correct then the period evolution of Cyg
X-3, determined by extreme mass-loss from the system (van Kerkwijk et
al. 1992; Kitamoto et al. 1995) will not be smooth, instead displaying
periods of accelerated lengthening during outbursts. The detection
and measurement of such `glitches' would be important both for
understanding the evolution of the Cyg X-3 system and estimating the
amount of additional circumstellar material present during outbursts.

The appearance and variability of the emission features in outburst is
suggestive of an asymmetric emitting region, and we propose that the
wind in Cyg X-3 is significantly flattened, probably in the plane of
the binary orbit. This may explain the intrinsic polarisation of the
infrared emission from Cyg X-3, which indicates a scattering plane
perpendicular to the radio jet axis.  The interpretation of a
flattened wind is supported by polarimetric and direct radio
interferometric observations revealing evidence for flattened winds in
other Wolf-Rayet stars.  A simple model for the V/R variability in
outburst, in the context of a flattened disc-wind, comprising a small
ionised zone around the compact object and continuous P-Cygni
absorption which erodes the blue-shifted wing, qualitatively explains
the observations.  Furthermore, a disc-like wind in the Cyg X-3 system
also naturally explains why we can have both a large infrared
luminosity and yet still observe the X-ray source, a problem
highlighted by Mitra (1996, 1998) as being very serious for a
spherically symmetric wind. While there is still significant
scattering of the X-rays along the line of sight (see Berger \& van
der Klis 1994) it will be considerably less than in the case of a
spherically symmetric wind. Additionally we note that the apparent
one-sidedness of the radio jet from Cyg X-3 in the latest VLBA
observations (Mioduszewski et al. 1998) may arise not from a jet
aligned near to the line of sight (implying a nearly face-on orbit
which is seemingly incompatible with the strong orbital modulations
observed) but instead from the obscuration of the receding (northerly)
jet by the far side of the disc-wind. This would naturally explain why
the jet is so apparently one-sided on small scales and yet symmetrical
on larger scales.

To conclude, the combination of a WR-like spectrum, high luminosity ($M_k
\leq -5$) and evidence for a disc-like wind supports the
interpretation of Cyg X-3 as a high-mass X-ray binary in a very
transient phase of its evolution.

\section*{Acknowledgements}

 We wish to thank George and Marcia Rieke for the use of their
near-infrared spectrometer and help during the observations. RPF would
like to thank Rudy Wijnands for help with the XTE ASM light curves,
and Marten van Kerkwijk, Elizabeth Waltman, Michiel van der Klis,
Simon Clark, Jan van Paradijs, Lex Kaper and Rens Waters for many
useful discussions. The MMT is jointly operated by the Smithsonian
Astrophysics Observatory and the University of Arizona. We thank the
staff at MRAO for maintenance and operation of the Ryle Telescope,
which is supported by the PPARC.  RPF was supported during the period
of this research initially by ASTRON grant 781-76-017, and
subsequently by EC Marie Curie Fellowship ERBFMBICT 972436.  MMH has
been supported by NASA through Hubble Fellowship grant
\#HF-1072.01-94A awarded by the Space Telescope Science Institute,
which is operated by the Association of Universities for Research in
Astronomy, Inc., for NASA under contract NAS 5-26555.

\appendix

\section{Observing logs}

Tables A1-4 below list the epochs and exposure times of every K- and
H-band spectum taken during all four observing runs. The reduced
spectra are downloadable from 

\smallskip

\begin{small}
ftp://cdsarc.u-strasbg.fr/pub/cats/J/MNRAS/vol/page
\end{small}

\smallskip

HJD, as used below, is heliocentric-corrected Julian Date, with
2450000.0 subtracted.  The `Exposures' column shows three numbers,
indicating the number of spectra averaged, the number of exposures and
the length of each exposure (in seconds).  Orbital phase at the {\em
start} of each observation is indicated by $\phi$.

We request that any use of these spectra in future publications makes
reference to this work. 

\begin{table} 
\caption{May/June 1996 observations}
\begin{tabular}{cccc}  
\hline
UT & HJD & Exposures & $\phi$ \\
29/05/1996 10:58:22    &   232.9577     &  4x30x4s     &  0.020  \\
29/05/1996 11:08:32    &   232.9648     &  4x30x4s     &  0.056  \\   
29/05/1996 11:18:05    &   232.9714     &  4x30x4s     &  0.089  \\
29/05/1996 11:37:20    &   232.9848     &  4x30x4s     &  0.156  \\
30/05/1996 11:13:02    &   233.9680     &  4x30x4s     &  0.079  \\
30/05/1996 11:22:24    &   233.9745     &  4x30x4s     &  0.112  \\
30/05/1996 11:32:06    &   233.9812     &  4x30x4s     &  0.146  \\
31/05/1996 11:11:55    &   234.9673     &  4x30x4s     &  0.083  \\
31/05/1996 11:22:01    &   234.9743     &  4x30x4s     &  0.119  \\
31/05/1996 11:31:48    &   234.9811     &  4x30x4s     &  0.153  \\
01/06/1996 11:13:12    &   235.9682     &  4x30x4s     &  0.096  \\
01/06/1996 11:24:06    &   235.9758     &  4x30x4s     &  0.134  \\
01/06/1996 11:34:17    &   235.9828     &  4x30x4s     &  0.169  \\
02/06/1996 06:49:07    &   236.7849     &  4x30x4s     &  0.185  \\
02/06/1996 06:59:10    &   236.7918     &  4x30x4s     &  0.220  \\
02/06/1996 07:11:39    &   236.8005     &  4x30x4s     &  0.264  \\
02/06/1996 07:21:32    &   236.8074     &  4x30x4s     &  0.298  \\
02/06/1996 07:36:23    &   236.8177     &  4x30x4s     &  0.350  \\
02/06/1996 07:46:54    &   236.8250     &  4x30x4s     &  0.386  \\
02/06/1996 07:57:09    &   236.8321     &  4x30x4s     &  0.422  \\
02/06/1996 08:12:41    &   236.8429     &  4x30x4s     &  0.476  \\
02/06/1996 08:25:56    &   236.8521     &  4x30x4s     &  0.522  \\
02/06/1996 08:36:32    &   236.8595     &  4x30x4s     &  0.559  \\
02/06/1996 08:49:25    &   236.8684     &  4x30x4s     &  0.604  \\
02/06/1996 09:00:13    &   236.8759     &  4x30x4s     &  0.641  \\
02/06/1996 09:10:51    &   236.8833     &  4x30x4s     &  0.678  \\
02/06/1996 09:24:01    &   236.8924     &  4x30x4s     &  0.724  \\
02/06/1996 09:34:00    &   236.8994     &  4x30x4s     &  0.759  \\
02/06/1996 09:43:42    &   236.9061     &  4x30x4s     &  0.793  \\
02/06/1996 09:58:03    &   236.9161     &  4x30x4s     &  0.843  \\   
02/06/1996 10:07:52    &   236.9229     &  4x30x4s     &  0.877  \\
02/06/1996 10:17:43    &   236.9297     &  4x30x4s     &  0.911  \\
02/06/1996 10:30:13    &   236.9384     &  4x30x4s     &  0.954  \\
02/06/1996 10:42:52    &   236.9472     &  4x30x4s     &  0.998  \\
02/06/1996 10:52:45    &   236.9541     &  4x30x4s     &  0.033  \\
02/06/1996 11:02:52    &   236.9611     &  4x30x4s     &  0.068  \\
02/06/1996 11:12:45    &   236.9679     &  4x30x4s     &  0.102  \\
02/06/1996 11:25:19    &   236.9767     &  4x30x4s     &  0.146  \\
02/06/1996 11:35:18    &   236.9836     &  4x30x4s     &  0.181  \\
03/06/1996 11:14:23    &   237.9691     &  4x30x4s     &  0.116  \\
03/06/1996 11:24:11    &   237.9759     &  4x30x4s     &  0.150  \\
04/06/1996 11:12:23    &   238.9678     &  4x30x4s     &  0.117  \\
04/06/1996 11:22:10    &   238.9746     &  4x30x4s     &  0.151  \\
04/06/1996 11:30:53    &   238.9806     &  4x30x4s     &  0.181  \\
05/06/1996 11:04:59    &   239.9627     &  4x30x4s     &  0.099  \\
05/06/1996 11:15:36    &   239.9701     &  4x30x4s     &  0.136  \\
05/06/1996 11:25:31    &   239.9770     &  4x30x4s     &  0.171  \\
06/06/1996 11:08:58    &   240.9656     &  4x30x4s     &  0.121  \\
06/06/1996 11:19:14    &   240.9727     &  4x30x4s     &  0.157  \\
06/06/1996 11:29:13    &   240.9796     &  4x30x4s     &  0.192  \\
07/06/1996 11:15:21    &   241.9700     &  4x30x4s     &  0.151  \\
07/06/1996 11:25:29    &   241.9770     &  4x30x4s     &  0.187  \\
07/06/1996 11:35:14    &   241.9838     &  4x30x4s     &  0.220  \\
08/06/1996 07:41:51    &   242.8218     &  4x30x4s     &  0.417  \\
08/06/1996 08:05:59    &   242.8385     &  4x30x4s     &  0.501  \\
08/06/1996 08:17:41    &   242.8467     &  4x30x4s     &  0.542  \\
\end{tabular}
\end{table}

\begin{table}
\caption{September 1996 observations}
\begin{tabular}{cccc}  
\hline\\
UT & HJD & Exposures & $\phi$ \\
22/09/1996 03:33:03    &   348.6507     &  4x30x4s     &   0.382  \\
22/09/1996 03:56:45    &   348.6671     &  4x30x4s     &   0.465  \\
22/09/1996 04:08:34    &   348.6753     &  4x30x4s     &   0.506  \\
22/09/1996 04:21:14    &   348.6841     &  4x30x4s     &   0.550  \\
22/09/1996 04:31:15    &   348.6911     &  4x30x4s     &   0.584  \\
22/09/1996 04:57:08    &   348.7091     &  4x30x4s     &   0.675  \\
22/09/1996 05:09:41    &   348.7178     &  4x30x4s     &   0.718  \\
22/09/1996 05:19:37    &   348.7247     &  4x30x4s     &   0.753  \\
22/09/1996 05:34:25    &   348.7349     &  4x30x4s     &   0.804  \\
22/09/1996 05:49:31    &   348.7454     &  4x30x4s     &   0.857  \\
22/09/1996 06:01:29    &   348.7537     &  4x30x4s     &   0.898  \\
22/09/1996 06:14:07    &   348.7625     &  4x30x4s     &   0.942  \\
22/09/1996 06:25:43    &   348.7706     &  4x30x4s     &   0.983  \\
22/09/1996 06:51:08    &   348.7882     &  4x30x4s     &   0.071  \\
22/09/1996 07:02:23    &   348.7960     &  4x30x4s     &   0.110  \\
22/09/1996 07:13:23    &   348.8037     &  4x30x4s     &   0.148  \\
22/09/1996 07:25:37    &   348.8122     &  4x30x4s     &   0.191  \\
22/09/1996 07:42:42    &   348.8240     &  4x30x4s     &   0.250  \\
22/09/1996 07:54:30    &   348.8322     &  4x30x4s     &   0.291  \\
22/09/1996 08:06:39    &   348.8407     &  4x30x4s     &   0.334  \\
22/09/1996 08:17:46    &   348.8484     &  4x30x4s     &   0.372  \\
22/09/1996 08:34:04    &   348.8597     &  4x30x4s     &   0.429  \\
23/09/1996 03:56:40    &   349.6670     &  4x60x4s     &   0.472  \\
23/09/1996 04:16:31    &   349.6808     &  4x60x4s     &   0.541  \\
23/09/1996 04:35:22    &   349.6939     &  4x60x4s     &   0.606  \\
23/09/1996 04:53:58    &   349.7068     &  4x60x4s     &   0.671  \\
\end{tabular}
\end{table}

\begin{table}
\caption{June 1997 observations}
\begin{tabular}{cccc}  
\hline\\
UT & HJD & Exposures & $\phi$ \\
16/06/1997 10:32:35    &   615.9407     &  4x60x4s     &   0.905  \\
16/06/1997 10:52:02    &   615.9542     &  4x60x4s     &   0.972  \\
17/06/1997 10:28:45    &   616.9381     &  4x60x4s     &   0.899  \\
17/06/1997 10:47:10    &   616.9509     &  4x60x4s     &   0.963  \\
17/06/1997 11:10:13    &   616.9669     &  4x60x4s     &   0.043  \\
18/06/1997 10:55:50    &   617.9570     &  4x60x4s     &   0.001  \\
18/06/1997 11:16:54    &   617.9716     &  4x60x4s     &   0.075  \\
18/06/1997 11:39:43    &   617.9875     &  4x60x4s     &   0.154  \\
19/06/1997 07:04:32    &   618.7964     &  4x60x4s     &   0.205  \\
19/06/1997 07:24:39    &   618.8104     &  4x60x4s     &   0.275  \\
19/06/1997 07:45:14    &   618.8247     &  4x60x4s     &   0.347  \\
19/06/1997 08:06:20    &   618.8393     &  4x60x4s     &   0.420  \\
19/06/1997 08:27:34    &   618.8541     &  4x60x4s     &   0.494  \\
19/06/1997 08:48:52    &   618.8688     &  4x60x4s     &   0.568  \\
19/06/1997 09:14:21    &   618.8865     &  4x60x4s     &   0.656  \\
19/06/1997 09:34:34    &   618.9006     &  4x60x4s     &   0.727  \\
19/06/1997 09:57:16    &   618.9163     &  4x60x4s     &   0.806  \\
19/06/1997 10:21:01    &   618.9328     &  4x60x4s     &   0.888  \\
19/06/1997 10:40:03    &   618.9461     &  4x60x4s     &   0.955  \\
19/06/1997 11:04:47    &   618.9632     &  4x60x4s     &   0.041  \\
19/06/1997 11:24:35    &   618.9770     &  4x60x4s     &   0.109  \\
20/06/1997 08:27:27    &   619.8540     &  4x60x4s     &   0.501  \\
20/06/1997 08:45:51    &   619.8668     &  4x60x4s     &   0.565  \\
20/06/1997 10:53:09    &   619.9552     &  4x60x4s     &   0.008  \\
20/06/1997 11:12:39    &   619.9687     &  4x60x4s     &   0.076  \\
\\
\end{tabular}
\end{table}

\begin{table}
\caption{October 1997 observations}
\begin{tabular}{cccc}  
\hline\\
UT & HJD & Exposures & $\phi$ \\
15/10/1997 04:25:20    &   736.6861     &  4x60x4s     &   0.567  \\
15/10/1997 04:54:05    &   736.7061     &  4x60x4s     &   0.667  \\
15/10/1997 05:14:48    &   736.7205     &  4x60x4s     &   0.739  \\
15/10/1997 05:36:05    &   736.7353     &  4x60x4s     &   0.813  \\ 
\end{tabular}
\end{table}

\end{document}